\documentclass[reprint,aps,prl]{revtex4-1}
\usepackage{graphicx}
\usepackage{bm}
\usepackage{color}
\usepackage{subfigure}
\usepackage{ulem}
\usepackage{amsmath}

\begin{document}
\title{The Chern Numbers of Interaction-stretched Monopoles in Spinor Bose Condensates}
\author{Tin-Lun Ho$^{\dagger, \ast}$, and Cheng Li$^{\dagger}$}
\affiliation{$^{\dagger}$ Department of Physics, The Ohio State University, Columbus, OH 43210, USA
\\
$^{\ast}$ Institute for Advanced Study, Tsinghua University, Beijing 100084, China} 
\date{\today}

\begin{abstract}
Using the Dirac and the Yang monopole in spinor condensates as examples, we show that  interactions can stretch the point singularity of a monopole into an extended manifold,  whose shape is strongly influenced by the sign of interaction. The singular manifold will cause the first and second Chern number to assume non-integer values when it intersects the surface on which the Chern numbers are calculated. This leads to a gradual decrease of the Chern numbers as the monopole moves away from the surface of integration, instead of the sudden jump characteristic of  a point monopole. A gradual change in $C_2$ has  in fact been observed in  the recent experiment by Spielman's group at NIST. By measuring the range of non-integer values of the Chern numbers as the monopole moves away from the surface of integration along different directions, one can map out the shape of the singular manifold  in the parameter space. 
\end{abstract}

\maketitle

Chern numbers appear in many areas of physics. They describe the structure of the manifold of quantum states, relating the local geometry to the global topology. Nontrivial topologies of quantum states can lead to a wide range of novel phenomena, ranging from monopoles and instantons in gauge theories\cite{Coleman,R,tHooft}  to the quantization of quantum Hall conductance in 2D electron gas\cite{QH}. In condensed matter, one often deals with multicomponent quantum states, parameterized by a set of parameters  ${\bf k}=(k_{1}, k_{2}, ..., k_{D})$. Depending on the number of components, as well as additional constraints, the topology of the manifold can be very complex. A topological structure that often emerges in these manifolds is the monopole. If the ground state is non-degenerate at each ${\bf k}$, then the monopole has a non-zero first Chern number $C_{1}$. If the ground state at each ${\bf k}$ is doubly degenerate, such as in the case of the the Yang monopole\cite{Yang} or electric quadrupole\cite{Mead}, then the monopole has non-zero second Chern number $C_{2}$\cite{C2a,C2b}. 

Recently, several groups have created a Dirac-like monopole using a two level system, and have verified $C_1 = 1$\cite{C1-1,C1-2}.
Very recently, Ian Spielman's group at NIST has succeeded in creating a Yang monopole in a five dimensional parameter space using a $^{87}Rb$ Bose spinor condensate that consists of four spin states\cite{Ian}. They have measured the non-abelian Berry curvature using the dynamic protocol of M. Kolodrubetz\cite{K}, and have demonstrated that the 2nd Chern number $C_2$ calculated over a four dimensional closed surface enclosing the monopole is an integer. Ideally, $C_2$ will jump from 1 to 0 as soon as the monopole leaves the 4D surface.  Experimentally, the change is found to be gradual.  
(Similar gradual change in $C_1$ has also been observed in Ref. \cite{C1-1,C1-2})
The authors have attributed this is to the difficulty in achieving adiabaticity in their measurements.  While non-adiabaticity can be a factor,  
we show here that in the case of Bose condensates, the gradual change will persist even when adiabaticity is achieved, for interaction will stretch out the discrete jump into a smooth descent. 
This is because as one approaches the monopole, the Zeeman-like energy caused by the  monopole field vanishes. For a spinor condensate, the interaction energy will always dominate over the ``Zeeman" energy within a finite volume surrounding the monopole. We shall show that within this volume,  the point monopole will be stretched out into an extended manifold of singularity. Consequently, the passage of the monopole through the 4D surface is no longer a sudden event, but a gradual process that extend over a distance in ${\bf k}$-space  that reflects the strength as well as the sign of the interaction.  When the singular manifold intersects with the 4D surface during the passage, the 2nd Chern number can assume non-integer values and will decrease in a continuous manner. 

Consider a Hilbert space where the state vectors are parametrized by a vector ${\bf k}$ in $D$ dimensional space. Let 
${\cal M}=\{ |a ({\bf k})\rangle \}$ be the subset of states of interest 
and  $\overline{\cal M} = \{|\overline{a} ({\bf k})\rangle\}$ be its complement.  
 The non-Abelian Berry connection $A_{\mu}^{ab}({\bf k})$  and curvature  $F_{\mu\nu}^{ab}({\bf k})$  in ${\cal M}$ are defined as 
\begin{eqnarray}
A_{\mu}^{ab}({\bf k}) &=& -i \langle a ({\bf k})| \partial_{\mu} | b({\bf k})\rangle,
\\
F_{\mu\nu}^{ab}({\bf k}) &=& \partial_{\mu} A_{\nu}^{ab} ({\bf k})-  \partial_{\nu} A_{\mu}^{ab} ({\bf k}) + i [ A_{\mu} ({\bf k}), A_{\nu} ({\bf k})]^{ab},\hspace{0.2in} 
\end{eqnarray}
where $\partial_{\mu}\equiv \partial/\partial k_{\mu}$. 
The 1st (or 2nd) Chern number can be defined on a closed 2 (or 4)
dimensional manifold $\cal K$ in the $\bf k$ space.
For the convinience, we could choose $\cal K$ as a spherical surface $S_2$ (or $S_4$). Explicitly, we have\cite{C2a,C2b}
\begin{eqnarray}
&C_1&=\frac{1}{2\pi}\int_{S^2} {\rm d}^{2}k\, \epsilon_{\mu\nu}{\rm Tr}F_{\mu\nu},\label{C1} \\
&C_2&=\frac{1}{32\pi^2}\int _{S^4} {\rm d}^{4}k\, \epsilon_{\mu\nu\rho\delta} \left({\rm Tr}\left[ F_{\mu\nu} F_{\rho\delta} \right]-{\rm Tr} F_{\mu\nu} {\rm Tr}F_{\rho\delta}\right)\label{C2},\hspace{0.2in} 
\end{eqnarray}
where the $\epsilon$'s are Levi-Civita symbols. 
For later discussions,  it is convenient to express the curvature tensor  { $\hat{F}_{\mu\nu}({\bf k}) = \sum_{a,b}  F_{\mu\nu}^{ab}({\bf k})  |a ({\bf k}) \rangle\langle  b({\bf k})|$
in terms of the projection operators of ${\cal M}$ and $\overline{\cal M}$, denoted as $P=\sum_{a} |a \rangle \langle  a |$ and $\overline{P}=\sum_{\overline{a}} |\overline{a} \rangle \langle\overline{a} |$ respectively. It is straightforward to show that \cite{P}
\begin{equation}
\hat{F}_{\mu\nu} =  -i  (\partial_{\mu} P) \overline{P} (\partial_{\nu} P) - ( \mu \leftrightarrow \nu) = -i P[\partial_{\mu}P, \partial_{\nu} P].
\label{P}
\end{equation}

{\bf {\em Interaction effects on a Dirac monopole:}} Consider  a two-component  spinor Bose condensate $\Psi= \sqrt{N}\Phi$ in a ``magnetic field" ${\bf b}$, where $N$ is the particle number and $\Phi^{\dagger}\Phi=1$.  It has the general form 
 $\Phi^{T}= (e^{-i \alpha/2}{\rm cos}\beta/2 ,  e^{i \alpha/2} {\rm sin}\beta/2 )$ up to a phase factor.  The spinor lies along $\bm{\hat{\ell}}$,   $\bm{\hat{\ell} \cdot \sigma}\Phi = + \Phi$, with 
 \begin{equation}
\bm{\ell} = {\rm cos}\beta\hat{\bf z} +  {\rm sin}\beta \left( {\rm cos}\alpha   \hat{\bf x} 
 + {\rm sin}\alpha   \hat{\bf y} \right) .
\label{ell}  \end{equation} 
To simplify the problem, we make the single-mode approximation, i.e. all bosons have the same spatial wavefunction. Within this approximation, the energy is $E =N({\cal E}_{Z} + {\cal U})$, where 
${\cal E}_{Z}$ is the ``Zeeman" energy and ${\cal U}$ is the interaction energy per particle, 
\begin{equation}
{\cal E}_{Z} = -  \Phi^{\dagger} {\bf k}\cdot  \bm{\sigma}  \Phi,  \,\,\,\,\,   {\cal U} = \frac{\gamma}{2}\left| \Phi^{\dagger} \sigma_{z} \Phi \right|^2, 
\label{M+u} \end{equation}
where $\bm{\sigma}$'s are  Pauli matrices,  and $\gamma$ is the interaction energy.  For a two-component Bose gas, the general form of the interaction energy is $(g_{1}N_{1}^2+ g_2 N_{2}^2+ 2g_{12} N_{1} N_{2})/V$. Considering the symmetric case, $g_1 =g_2 = \bar g$, the interaction energy becomes  that in Eq.(\ref{M+u}) with $\gamma=(\bar{g}-g_{12})N/2V$,  plus a term depending on the total particle number $N$, which can then be ignored. 
The energy per particle is
\begin{equation}
E/N = -k (\bm{ \hat{k} \cdot \hat{\ell} })  + \gamma ( \hat{\bf z} \bm{ \cdot \hat{\ell} })^2/2.
\label{E1} \end{equation}
When $\gamma=0$, we have $\hat{ \bm{\ell}} = \hat{\bf k}$, and the ground state is a monopole at the origin of ${\bf k}$-space. 
We shall refer to this as a Dirac monopole\cite{Dirac} as they have the same Berry connection\cite{Berry}.
When $\gamma \neq 0$,  $\hat{ \bm{\ell}}$ no long aligns with $\hat{\bf k}$. However, it 
remains cylindrical symmetric about the $\hat{\bf z}$-axis and mirror symmetric about the $x$-$y$ plane. 
Precisely, it means
\begin{equation}
\alpha=\phi, \,\,\,\,\, \beta(k, \pi-\theta) = \pi-\beta(k, \theta), 
\label{sym} \end{equation}
where $(k, \theta, \phi)$ is the polar coordinate of ${\bf k}$, and $\beta= \beta(k, \theta)$ is given by the stationary condition of  Eq.(\ref{E1}),   
\begin{equation}
\begin{split}
k  \sin(\beta - \theta) - (\gamma/2){\rm sin}2\beta = 0,\\
k \cos(\beta-\theta)-\gamma\cos2\beta > 0.
\end{split}\label{stationary}
\end{equation}

{\bf {\em Repulsive interaction $\gamma>0$:}}  In this case, interaction favors $\bm{\hat{\ell}}$  to lie in the $x$-$y$ plane. Consequently, it stretches out the original monopole into a singular line segment $|k_{z}| < \gamma$ along the $\hat{\bf z}$ axis as shown in  Fig.\ref{fig:sing_gamma>0}. It follows from Eq.(\ref{stationary})  that
\begin{equation}
{\rm cos}\beta (k, 0)  = k/\gamma \,\,\, ({\rm or} \,\,\,  1) \,\,\,\,\,\,   {\rm for } \,\,\,\,\,   k < \gamma \,\,\, ({\rm or} \,\,\,  k>\gamma). 
\label{cos}  \end{equation}
The spin texture is discontinuous on this line segment.  
Substituting the projection operator of the ground state, $P=(1+ \hat{\bm{\ell}}\cdot\bm{\sigma})/2$,  into 
Eq.(\ref{C1}) and (\ref{P}), we have 
\begin{equation}
C_{1} = \frac{1}{4\pi} \int_{S^2}  {\rm d} \theta {\rm d} \phi  \,\, \,
\hat{\bm{\ell }} \cdot \partial_{\theta}  \hat{\bm{\ell }} \times  \partial_{\phi} \hat{\bm{\ell }} 
=\int_{S^2} \, {\rm sin}\beta \,\, \frac{ \partial\beta }{\partial \theta} \,  \frac{{\rm d} \theta }{2},
\label{C1U1} \end{equation}
where $\beta=\beta(k, \theta)$. If $S^2$ is a sphere with its center shifted down from the monopole by distance $D$ along  the $\hat{\bf z}$-axis, then $(k,\theta)$ in Eq.(\ref{C1U1}) is related through $({\bf k}+D\hat{\bf z})^2=R^2$, i.e. $k^2 + 2kD {\rm cos}\theta + D^2 -R^2 =0$. If the monopole is inside $S^2$, $D<R$ (see Fig.\ref{fig:inside}), we have
\begin{equation}
C_1 = \left[ {\rm cos}\beta(R-D, 0)- {\rm cos}\beta(R+D,\pi)\right]/2, \,\,\,\,\, D<R.
 \end{equation}
When the original point monopole is outside the sphere $S^2$, $D>R$ (see Fig.\ref{fig:outside}).
 $\theta$ starts from $\pi$ (with $k=D-R$), then decrease to 
$\theta^{\ast} = \pi-\arccos\frac{\sqrt{D^2-R^2}}{D} > \pi/2$, and then increases back to $\pi$ (with $k=R+D$). 
As a result, we have  
\begin{equation}
C_1 = \left[ {\rm cos}\beta(D-R, \pi)- {\rm cos}\beta(R+D,\pi)\right]/2,  
\,\,\,\,\, D>R. 
 \end{equation}
As $S^2$ moves away the from the monopole, $C_1$ drops from 1 to 0 continuously over the range of $D$ where $S^2$ intersects the singular line segment. This is 
shown in Fig.\ref{fig:C1}, where we have chosen a sphere $S^2$ with radius $R>\gamma$. 

{\bf {\em Attractive interaction $\gamma<0$:}}
In this case, interaction favors $\bm{\hat{\ell}}$ to lie along $\pm\hat{\bf z}$. 
It therefore stretches the  monopole into a singular disc  of radius  $k=|\gamma|$ in the $x$-$y$-plane as shown in
 Fig.\ref{fig:sing_gamma<0}.  
It follows from Eq.(\ref{stationary}) that on the disc,  
\begin{equation}
{\rm sin}\beta (k, \pi/2)  = k/\gamma  \,\,\, ({\rm or} \,\,\, 1 ) \,\,\,\,\,\,   {\rm for } \,\,\,\,\,   k < |\gamma| \,\,\, ({\rm or} \,\,\,  k>|\gamma|). 
\label{sin} \end{equation} 
The decrease of  $C_{1}$ can be calculated in similar way as the repulsive case.(See Supplementary Material.) The behavior of $C_{1}$ as 
 the monopole leaves the surface $S^{2}$ is shown in Fig.\ref{fig:C1}.  It shows that $C_{1}$ decreases faster than the repulsive case. This is because the displacement vector $D\hat{\bf z}$ is normal to 
 the singular disc. Consequently, the range of $D$ where $S^2$ intersects the singular manifold is  shorter. (See Fig.\ref{fig:sing_gamma<0}). 
 
\begin{figure}[htbp]
\centering
\subfigure[$\gamma>0$]
{
\label{fig:sing_gamma>0}
\includegraphics[width=3in]{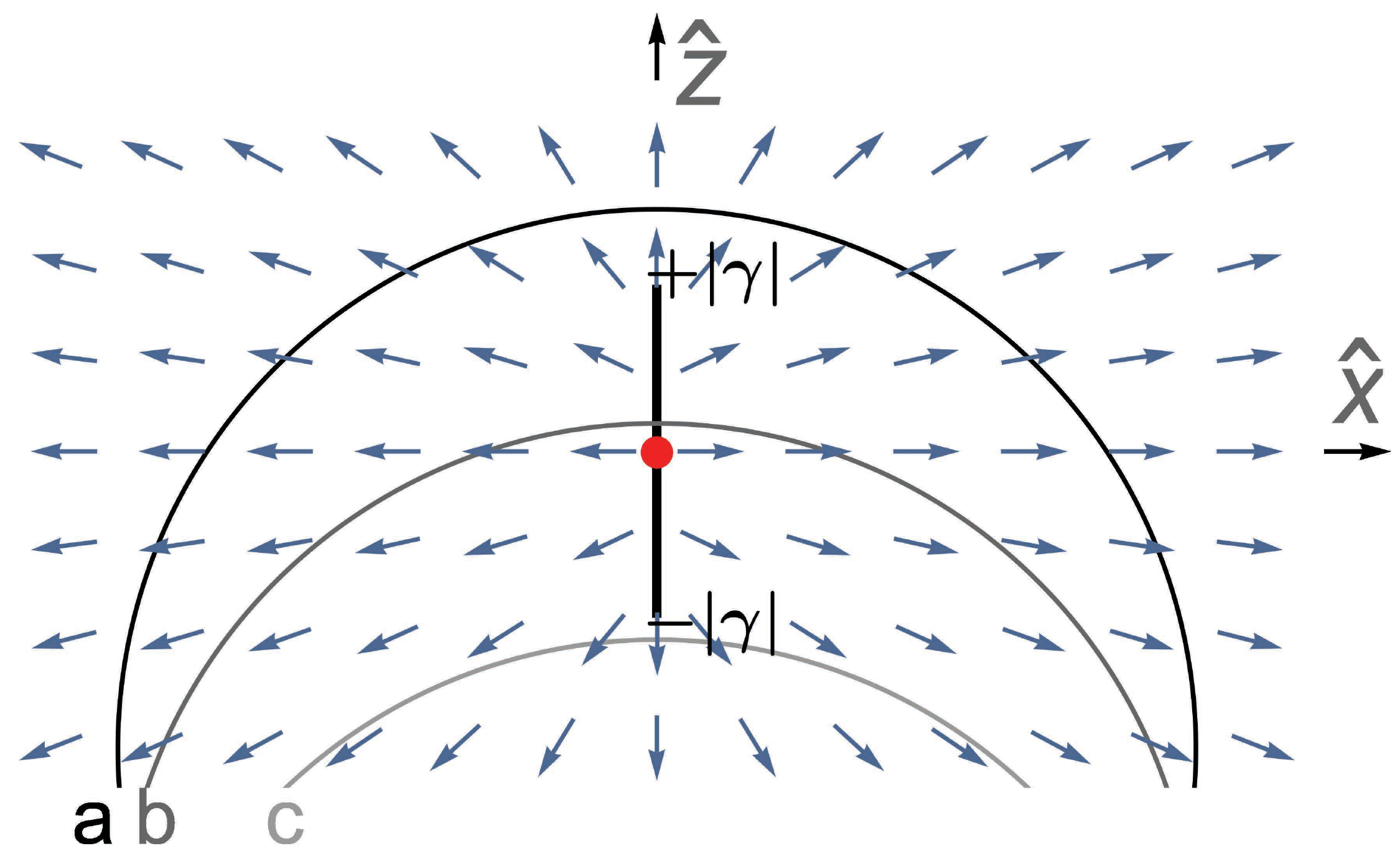}
}
\subfigure[$\gamma<0$]
{
\label{fig:sing_gamma<0}
\includegraphics[width=3in]{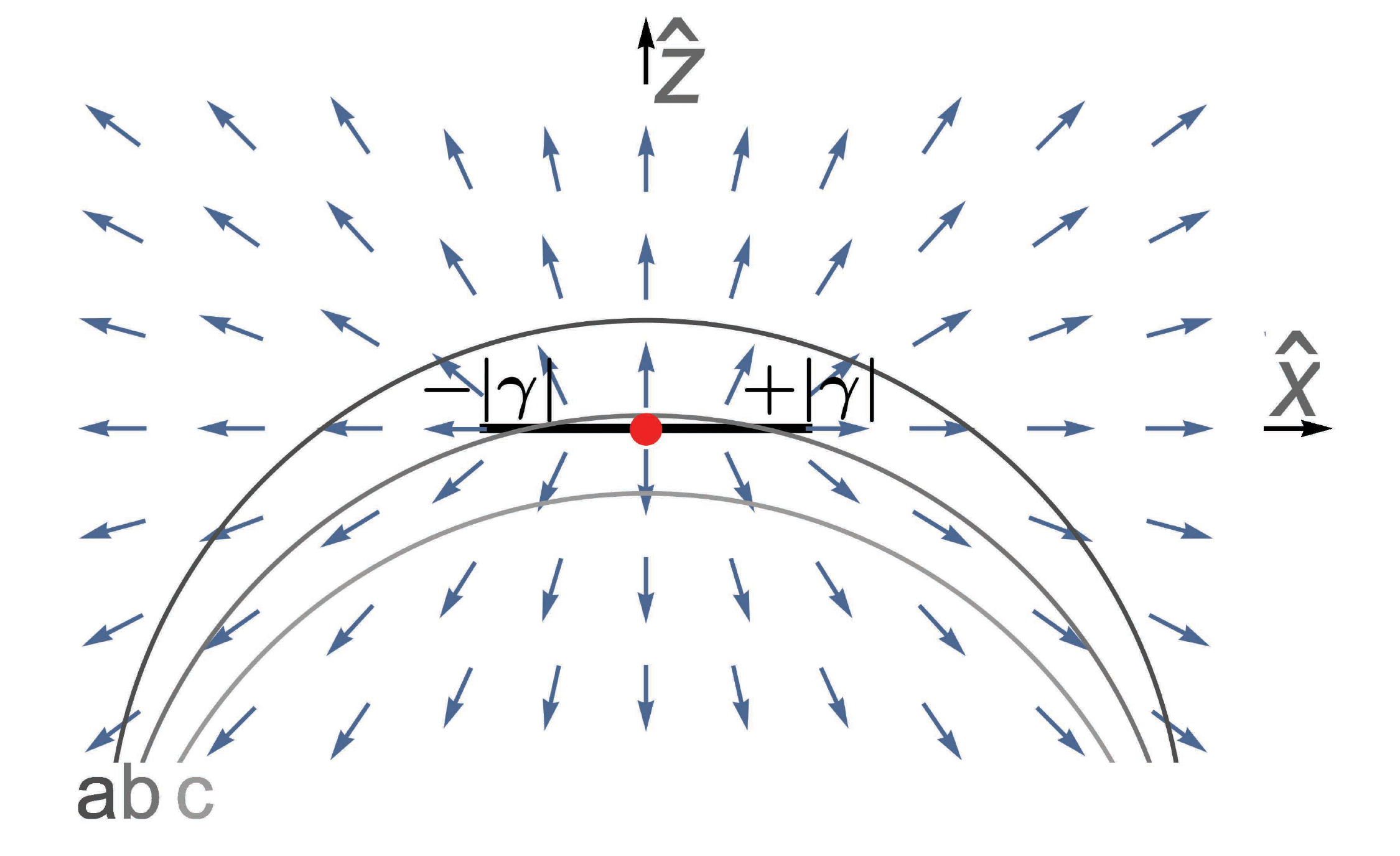}
}
\caption{The spin texture $\bm {\ell}({\bf k})$ in parameter space around the original monopole at ${\bf k}=0$  for 
repulsive interaction $\gamma>0$ and attractive interaction $\gamma<0$ respectively. In both cases, the spin texture (represented by the arrows) is cylindrical symmetric about the $\hat{\bf z}$ axis and has mirror symmetry about the  $x$-$y$ plane.  Repulsive  interaction stretches the point monopole at ${\bf k}=0$  into a singular line segment of length $2\gamma$ along the $\hat{\bf z}$ axis (Fig. \ref{fig:sing_gamma>0}). The singular line segment is characterized by a non-zero angle $\beta(k, 0)$ for all $\phi$, (Eq.(\ref{cos})). Attractive  interaction stretches the point monopole into a disc of singularity of radius $|\gamma|$ in the $x$-$y$ plane (Fig. \ref{fig:sing_gamma<0}).  The singularity is characterized by a non-zero  deviation of $ \beta (k , \pi/2)$ from $\pi/2$, (Eq.(\ref{sin})). The sequence of circles represent the surfaces of integration $S^2$ shifted down for the monopole by different displacements $-D\hat{\bf z}$.   $C_1$ is 1 (and 0) for surface $a$ (and $c$) that encloses (or excludes) the singular line (or disc), and assumes non-integral  values for surface $b$ that intersects the singular manifold. The change of $C_{1}$ as a function of $D$ is shown in Fig \ref{fig:C1}. }
\label{fig:sing}
\end{figure}

\begin{figure}[htbp]
\centering
\subfigure[$D<R$]
{
\label{fig:inside}
\includegraphics[height=1.35in]{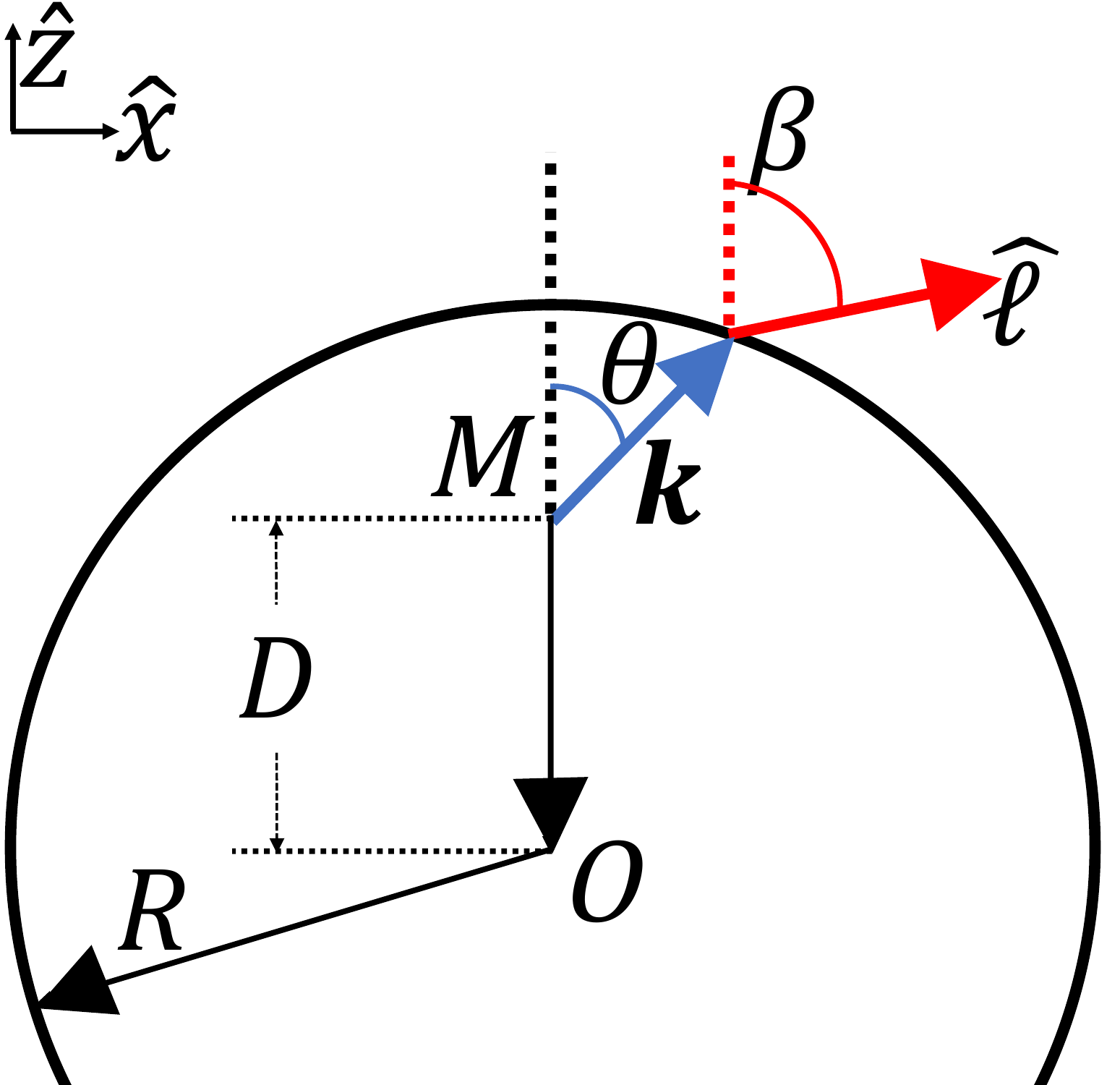}
}
\subfigure[$D>R$]
{
\label{fig:outside}
\includegraphics[height=1.35in]{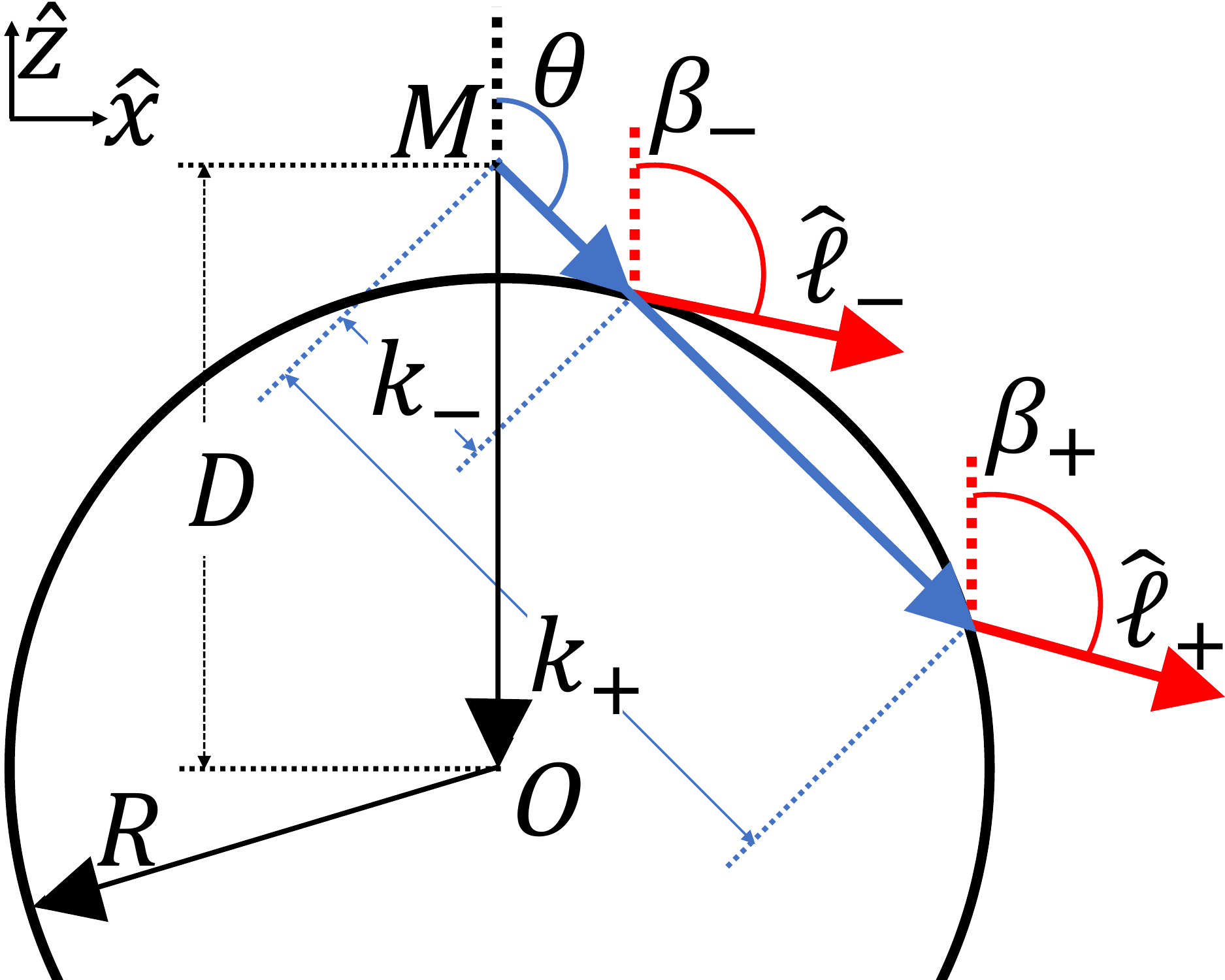}
}
\caption{The integration sphere $S^2$ (centered at $O$) is displaced from the Dirac monopole $M$ by $- D\hat{\bf z}$.  $S^2$ is denoted by the vector ${\bf k}$ when measured at $M$, with $({\bf k}+D\hat{\bf z})^2=R^2)$. The spin vector $\hat{\bm \ell}$ will not be parallel to $\hat{\bf k}$ unless  interaction $\gamma = 0$. $\theta$ and $\beta$ are the polar angles of $\hat{\bf k}$ and $\hat{\bm \ell}$.  
When $M$ is inside $S^2$, $D<R$ (see Fig.\ref{fig:inside}), $\theta$ increases from 0 to $\pi$ as the sphere is traversed from the north pole to the south pole.  
When $M$ is outside $S^2$, $D>R$, (see Fig.\ref{fig:outside}),  $\theta$ starts from $\pi$, then reaches a minimum angle $\theta^{\ast}$ (which is greater than $\pi/2$), and then back to $\pi$. For every angle $\theta$, the surfaces is given by two $k$ values: a branch $k_{-}$ when $\theta$ decreases from $\pi$ to $\theta^{\ast}$, and a branch $k_{+}$ for $\theta$ increases from $\theta^{\ast}$ to $\pi$.
The same diagram applies to the Yang monopole. In that case, $S^2$ becomes $S^4$. The directions $\hat{\bf x}$ and $\hat{\bf z}$ become $\hat{\bf e}_{1}$ and $\hat{\bf e}_{5}$.  These figures do not display the polar angles of $\hat{\bf n}$. They are meant to be schematic representations of a cross section of $S^4$ for given ${\bf n}$.  }
\label{fig:integ}
\end{figure}

\begin{figure}[htbp]
\centering
\subfigure[$C_1$]
{
\label{fig:C1}
\includegraphics[width=2in]{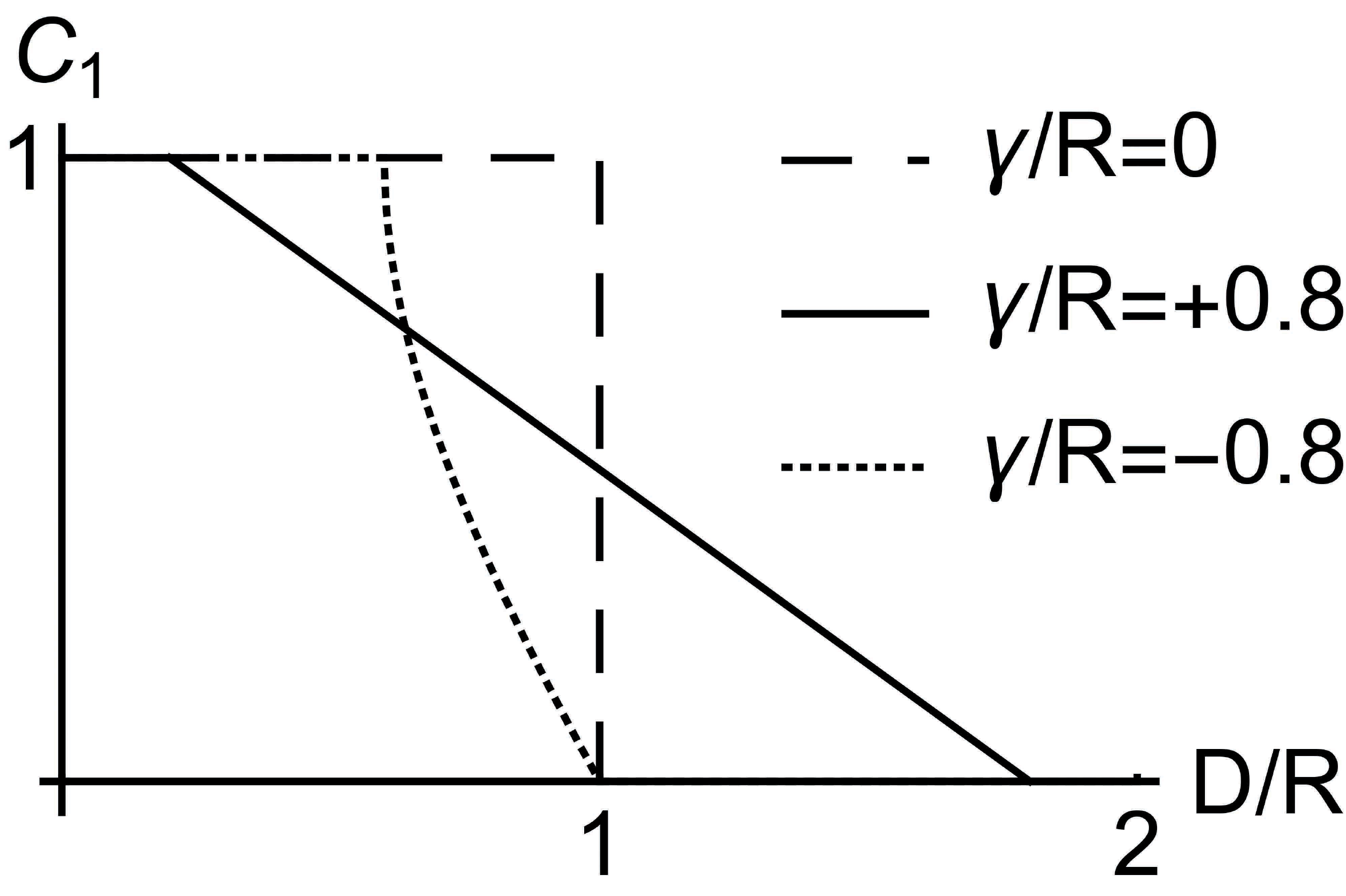}
}
\subfigure[$C_2$]
{
\label{fig:C2}
\includegraphics[width=2in]{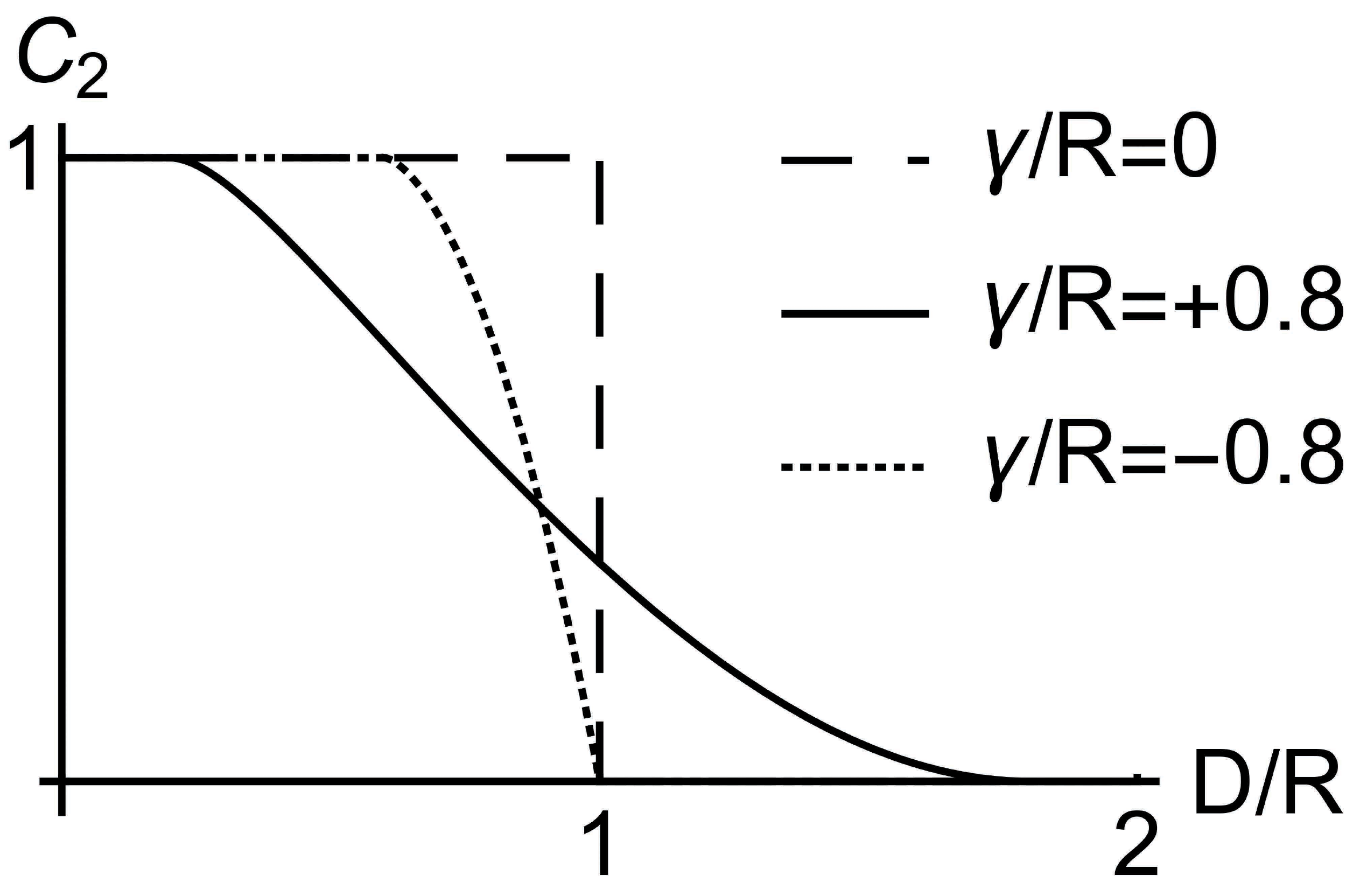}
}
\caption{$C_{1}$ ($C_{2}$) as a function of the distance $D$ between the monopole and 
the center of the integration sphere $S^2$ ($S^4$) for the cases (i) $\gamma=0$,  (ii) $\gamma = + 0.8R$, (iii) $\gamma = - 0.8R$.  In case  (i), 
 $C_{1}$  ($C_{2}$) drops abruptly from 1 to 0 as soon as the monopole leaves $S^2$ ($S^{4}$).  When $\gamma\neq 0$, the decrease is gradual.  The interval of the decrease reflects the range of the displacement where the integration surface intersects the singular manifold.  This explains the different width of decrease between  (ii) and (iii), as  interactions of the same strength $|\gamma|$ but different signs lead to singular manifolds of different shapes and orientation. }
\label{fig:C-D}
\end{figure}

{\bf {\em Yang monopole:}}  The Yang monopole\cite{Yang} engineered in Ref.\cite{Ian} is made of a spinor condensate with four spin states of $^{87}Rb$, with a ``Zeeman" energy ${\cal E}_{Z}= \Phi^{\ast} h_{Z} \Phi$  in a 5D parameter space, 
\begin{equation}
h_{Z} =  - \sum_{\alpha=1}^{5} b_{\alpha}  \Gamma_{\alpha} = - {\bf b}\cdot \bm{\Gamma},
\label{5Z}
\end{equation}
where $\Gamma_{\alpha}$ are the $4\times4$ Gamma matrices.
We shall represent the Gamma matrices as direct products of two sets of Pauli matrices $\{ \bm{\tau} \}$ and $\{ \bm{\sigma}  \}$.  If $\Phi$ is regarded as a spin-3/2 particle,  ($J_{z}= \sigma_{z}/2 + \tau_{z}$, $J_{+}= \sqrt{3} \sigma_{+} + 2\tau_{+}\otimes\sigma_{-}$), the general form of the Hamiltonian with time reversal symmetry 
up to a unitary transformation is of the quadrupolar form $Q_{ab}J_{a}J_{b}$, and can be written as 
\begin{equation}
h_{Z}=  - k_{z}\tau_{z}\otimes \hat{\bf n}\cdot \bm{\sigma}  - k_{x}\tau_x -k_{y}\tau_{y},  
\label{Ho}
\end{equation}
where $\hat{\bf n}$ is a unit vector. (See Supplementary Material)
This means that in Eq.(\ref{5Z}) we have chosen
\begin{eqnarray}
(\Gamma_{1},  \Gamma_{2}, \Gamma_{3},  \Gamma_{4}, \Gamma_{5}) &=& 
(\tau_x, \tau_y, \tau_z \sigma_x,  \tau_z \sigma_y,  \tau_z \sigma_z),
\label{0.2in} \\
 (b_{1},   b_{2}, b_{3},  b_{4}, b_{5}) &=& (k_x, k_y, k_z  {n}_{x},  k_z  {n}_{y},  k_z  {n}_{z}); \,\, 
 \end{eqnarray} 
hence ${\bf b}^2 ={\bf k}^2$. The Hamiltonian Eq.(\ref{Ho}) differs from that in Ref.\cite{Ian} by a simple unitary transformation. 

Eq.(\ref{Ho}) has two degenerate ground states. They are 
 \begin{eqnarray}
\Phi^{(1)}=  \zeta(\hat{\bf k})\otimes\chi (\hat{\bf n}),&\,\,\,&   \Phi^{(2)} =  {\cal T}\Psi \equiv  i \tau_{x} \otimes \sigma_{y}\Psi^{(1)\ast}; \hspace{0.2in} \label{Phi12}   \\
\hat{\bf n}\cdot \bm{\sigma}\chi (\hat{\bf n})= + \chi(\hat{\bf n}),& \,\,\, & \hat{\bf k}\cdot \bm{\tau}\zeta(\hat{\bf k}) = + \zeta(\hat{\bf k}); 
\hspace{0.5in} \label{xxx}  \end{eqnarray}
where ${\cal T}=   i \tau_{x}\otimes \sigma_{y} K$ is the time reversal operator,  $K$ is the complex conjugation,   $\chi(\hat{\bf n})$ and $\zeta(\hat{\bf k})$ are spin eigenstates in the $\bm{\sigma}$-space and $\bm{\tau}$-space
along $\hat{\bf n}$ and $\hat{\bf k}$ respectively, and ${\bf k} =(k_x, k_y, k_z)= k\hat{\bf k}$. 
Both states have  the same (unit) ``spin vector"
$\langle \Gamma_{\alpha} \rangle = \hat{b}_{\alpha}$, forming a monopole at the origin of the 5D ${\bf b}$-space. 
This monopole can also be viewed as a 3D monopole at the origin of ${\bf k}$-space (or in ${\bf k}'$ -space, ${\bf k'}\equiv (k_x, k_y, -k_z)$) with a 2-sphere $S^{2}_{\hat{\bf n}}$ ( or $S^{2}_{- \hat{\bf n}}$) attached to each ${\bf k}$ (or ${\bf k}'$) point.  Because of time reversal symmetry, $C_1=0$ for the Yang monopole.

{\bf {\em Interaction effects on a Yang Monopole:}}
The general form of interaction of a four state system is quite involved, as it will contain 
 density-density, ``spin-spin", and ``spin-exchange" interactions. Rather than studying the general case, we  consider a simple model to make the physics and the calculation transparent. 
 The model interaction energy (per particle) we consider is 
\begin{equation}
{\cal U}= -\gamma \left(  \Phi^{\ast}\Gamma_{3} \Gamma_{4} \Gamma_{5}\Phi\right)^2 /2= \gamma (\Phi^{\ast}\tau_{z} \Phi)^2/2. 
\end{equation} 
Since this ${\cal U}$ also preserves  time reversal symmetry, the eigenstates state of the interacting gas remain doubly degenerate,  hence $C_1=0$.  It is easy to see that the ground states  still has the form Eq.(\ref{Phi12}) with the same spinor $\chi$  in $\bm{\sigma}$-space but with the spinor $\zeta$ in $\bm{\tau}$-space pointing along a direction $\hat{\bm{\ell}}$ that is  different from $\hat{\bf k}$. The direction of $\bm \ell$ is determined by its energy, which is again given by Eq.(\ref{E1}).
 Consequently, $\hat{\bm{\ell}}$ is also given by Eq.(\ref{stationary}). 
The texture  $\hat{\bm{\ell}}$ is again represented by 
 Fig.\ref{fig:sing}, with $\hat{\bf x}\rightarrow \hat{\bf e}_{1}$, 
 $\hat{\bf z}\rightarrow \hat{\bf e}_{5}$, and with circles  representing $S^4$. 

Despite the change of spinor $\zeta$ in $\bm \tau$-space, the two degenerate states $\Phi^{(i)}$ in 5D space still have the same spin vector, given by $\langle \Gamma_{\alpha} \rangle_{\Phi^{(i)}} =\hat{m}_{\alpha}$,
$i=1,2$, where  
\begin{equation}
(\hat{m}_{1},  \hat{m}_{2}, \hat{m}_{3},  \hat{m}_{4}, \hat{m}_{5}) = (\hat{\ell}_x, \hat{\ell}_y, \hat{\ell}_z  \hat{n}_{x},  \hat{\ell}_z  \hat{n}_{y},  \hat{\ell}_z  \hat{n}_{z}), 
\end{equation}
and we have $\hat{\bf m}\cdot\bm{\Gamma}\Phi^{(i)} = + \Phi^{(i)}$.  It is also easy to show that the two degenerate excited states  $ \Phi^{(3)}$ and  $ \Phi^{(4)}$ satisfy $\hat{\bf m}\cdot\bm{\Gamma}\Phi^{(i)} = - \Phi^{(i)}$, $i=3,4$. So the 
 projection operator for the  degenerate ground state manifold is $P = (1+  \hat{\bf m}\cdot \bm{\Gamma})/2$.  Then Eq.(\ref{P})  implies 
\begin{equation}
C_{2} = \frac{1}{64\pi^2} \int_{S^{4}} \hat{m}^{a} \partial_{\mu_1}\hat{m}^{b}  \partial_{\mu_2} \hat{m}^{c} 
 \partial_{\mu_3} \hat{m}^{d}  \partial_{\mu_4} \hat{m}^{e}  \epsilon_{\mu_1\mu_2\mu_3\mu_4} \epsilon^{abcde}.
\label{C2A} \end{equation}
(See also Ref.\cite{C2-spin, C2-spin2}).  If we denote $S^{4}_{\hat{\bf m}}$ be the 4-sphere  representing all possible orientations $\hat{\bf m}$ in the 5D space, then Eq.(\ref{C2A}) counts the number of times the  surface $S^{4}$  wraps around $S^{4}_{\hat{\bf m}}$. 
It is an integer if 
$\hat{\bf m}$ is smooth  on  $S^{4}$.  On the other hand, when 
$S^{4}$ intersects with the singular manifold, 
$\hat{\bf m}$ has singularities on the integration surface, and $C_2$ will not be an integer.

We now  calculate $C_2$ on a 4D spherical of radius $R$ shifted down from the monopole along $\hat{\bf e}_{5}$ direction. 
Let $(\theta, \phi)$, $(\theta', \phi')$ 
 be the polar angles of $\hat{\bf k}$ and $\hat{\bf n}$ respectively. The equation for the integration surface $S^{4}$ is  $( {\bf b}_{5} + D\hat{\bf e}_{5})^2= R^2$, or 
\begin{equation} 
D^2 + k^2 + 2kD {\rm cos}\theta {\rm cos}\theta'=R^2.
\label{4D} \end{equation}
See Fig.\ref{fig:integ}. Expressing $\hat{\bf m}$ in Eq.(\ref{C2A}) in terms of 
the polar angles $(\beta, \alpha=\phi)$ and $(\theta', \phi')$ of $\hat{\bm \ell}$ and $\hat{\bf n}$, 
Eq.(\ref{C2A}) then becomes 
{\begin{equation}
C_{2} = \frac{3}{2} \iint_{S^{4}} {\rm d}\theta {\rm d}\theta' \,\, {\rm cos}^2\beta {\rm sin}\beta \frac{{\rm d}\beta}{{\rm d}\theta} {\rm sin}\theta'. 
\label{C2B} \end{equation}


The angle $\beta$ is a function of $k$, and $\theta$, given by Eq.(\ref{stationary}).  Since $k$ and $\theta$ are related to $\theta'$ through Eq.(\ref{4D}), the integrals in Eq.(\ref{C2B}) can not be performed independently. 
 While Eq.(\ref{C2B}) can be evaluated analytically, the calculation is much more involved than  that for $C_1$ as the geometry of 
$S^4$ (Eq. (\ref{4D}}) is more complex than that of $S^{2}$ (where $\theta'$ is absent).  
The details of the calculation is presented in Supplementary Material.  Its behavior is similar to  $C_{1}$ for the Dirac monopole. That is to say, 
 $C_2$ decreases from 1 to 0 continuously as the monopole moves away from the integration surface $S^4$. The range of displacement $D$ where 
 the decrease takes place reflects is the interval where $S^4$ intersects the singular manifold. See Fig.\ref{fig:C2}.
 
 
 
{\bf {\em Final remarks:}}  We have shown that even with a very simple form of the interaction, a point monopole (be it of Dirac or Yang type) will be stretched into an extended manifold of singularities.  
  The extended nature of this singular manifold is the origin of the non-integer Chern numbers. It occurs when the integration surface that intersects with singular manifold. Including more interaction parameters will further expand this singular manifold. The continuous drop of Chern numbers as the monopole moves away to infinity is therefore an intrinsic property of an interacting systems. In principle, the shape of this manifold can be revealed by measuring the size of interval where the Chern numbers drop
 from 1 to 0 as the monopole moves away from the integration sphere  along different directions in  parameter space. 
 The gradual decrease of $C_2$ has in fact showed up in the data of Ref.\cite{Ian}. While non-adiabaticity can be a factor,  it has an intrinsic contribution from interaction.

 This work is supported by the NSF Grant DMR-0907366, the MURI Grant FP054294-D, and the NASA Grant on Fundamental Physics 1541824.

\end{document}


\title{Supplementary Material: The Chern Numbers of Interaction-stretched Monopoles in Spinor Bose Condensates}
\author{Tin-Lun Ho, and Cheng Li}
\date{\today}
\maketitle
\renewcommand\theequation{S\arabic{equation}}
\renewcommand{\thefigure}{S\arabic{figure}}
\renewcommand{\thetable}{S\arabic{table}}

\section{Time reversal invariant Hamiltonian of a spin 3/2 particle:}

We shall represent the spin operators of a spin 3/2 particle, ($4\times 4$ matrices), as products of two two-level systems ($\bm \sigma$ and $\bm \tau$), 
$J_{+} = \sqrt{3}\sigma_{+} + \tau_{+}\otimes\sigma_{-}$, $J_{z} = \frac{1}{2} \sigma_{z} + \tau_{z}$. 
Under time reversal, ${\bf J}\rightarrow - {\bf J}$. This means $\bm \sigma \rightarrow 
- \bm \sigma$; $\tau_{z} \rightarrow - \tau_{z}$,  $\tau_{x} \rightarrow + \tau_{x}$, 
$\tau_{y} \rightarrow + \tau_{y}$.  The time reversal operator is then 
${\cal T}= i \tau_{x}\otimes \sigma_{y} K$, where $K$ is complex conjugation. 
The Hamiltonian of a spin-3/2 particle has the general form
$H= A {\mathbb 1_4}+  \sum_j B_j  \tau_j + \sum_i C_i \sigma_i +  \sum_{ij} D_{ij}   \tau_{j}\otimes \sigma_{i}$, 
where $A, B_j, C_i$, and $D_{ij}$ are all real.  For $H$ to be time reversal invariant, we need $C_i= B_{z} =D_{ix}=D_{iy}=0$ for all $i=x,y,z$, resulting in 
\begin{equation}
H= k_{x}\tau_{x} + k_{y}\tau_{y} + k_{z}\tau_{z}\otimes\hat{\bf n}\cdot \bm \sigma ,
\label{HH}\end{equation}
where  $k_{x}=B_{x}$, $k_{y}=B_{y}$, $k_{z}n_{i} = D_{iz}$, and we have set $A=0$ without loss of generality. 
 Since one can choose a different  representation of ${\bf J}$ in terms of $\bm \sigma$ and $\bm \tau$ by applying a unitary transformation. Eq.(\ref{HH}) is the general time invariant form up to a unitary transformation. 

Next, we note that Eq.(\ref{HH}) is equivalent to the quadrupolar form,  
$H=Q_{ij}J_{i}J_{j}$, where $Q$ is a symmetric traceless matrix.  To see this, we write 
$H= aJ_{x}^2 + b J_{y}^2 - (a+b) J_{z}^2 +c(J_{x}J_{y} + J_{y}J_{x}) + d(J_{x}J_{z} + J_{z}J_{x}) + e(J_{y}J_{z} + J_{z}J_{y})$, which can be written as 
\begin{equation}
H = \sqrt{3} \left(\begin{array}{cc} {\cal A}\sigma_{z} + {\cal D}\sigma_{x} + {\cal E}\sigma_{y} &  
 {\cal B} -i{\cal C}  \\
{\cal B}+i{\cal C}   &  - ({\cal A}\sigma_{z} + {\cal D}\sigma_{x} + {\cal E}\sigma_{y})
\end{array} \right), 
\label{HHH} \end{equation}
where ${\cal A}= (-\sqrt{3}/2)( a+b)$, ${\cal B}=(a-b)/2$, $C=c$, ${\cal D}=d$, ${\cal E}=e$. This is of the form 
Eq.(\ref{HH}) with ${\cal B}=k_{x}$, ${\cal C}=k_{y}$, ${\cal D}=k_{z}n_{x}$, 
${\cal E}=k_{z}n_{y}$, ${\cal A}=k_{z}n_{z}$. 
\section{The 1st Chern number of the stretched Dirac monopole}
\subsection{Under repulsive interaction $\gamma>0$}
The method to calculate $C_1$ has already given by \textcolor{black}{Eq.(11)(13)(14)} in the text.
Based on these equations, we could got
\begin{equation}
C_1(\gamma>0,D<R) = \left\{
        \begin{array}{cc}
          1 & 0<\gamma<R-D \\
          \frac{R-D+\gamma}{2\gamma} & R-D<\gamma<R+D \\
          \frac{R}{\gamma} & \gamma>R+D\\
        \end{array}
      \right.,
\label{eqn:C11}
\end{equation}
\begin{equation}
C_1(\gamma>0,D>R) = \left\{
        \begin{array}{cc}
          0 & 0<\gamma<D-R \\
          \frac{R-D+\gamma}{2\gamma} & D-R<\gamma<R+D \\
          \frac{R}{\gamma} & \gamma>R+D\\
        \end{array}
      \right..
\label{eqn:C12}
\end{equation}

Table \ref{tab:0-1} and Table \ref{tab:0-2} show the details about $\beta$ and $C_1$ when the monopole is inside and outside the sphere respectively.

\begin{table}[htbp]
\centering
\caption{$C_1$ under repulsive interaction $\gamma>0$, when the monopole is inside the sphere $D<R$}
\label{tab:0-1}
\begin{tabular}{|c|c|c|c|}
\hline
$\gamma$ & $\cos\beta(R-D,0)$  &$\cos\beta(R+D,\pi)$ & $C_1$\\ \hline
$0<\gamma<R-D$         &$1$ &$-1$&$1$                                  \\ \hline
$R-D<\gamma<R+D$ & $(R-D)/\gamma$ & $-1$ & $(R+\gamma-D)/2\gamma$\\ \hline
$\gamma>R+D$ &$(R-D)/\gamma$ & $-(R+D)/\gamma$ &$R/\gamma$\\ \hline
\end{tabular}
\end{table}

\begin{table}[htbp]
\centering
\caption{$C_1$ under repulsive interaction $\gamma>0$, when the monopole is outside the sphere $D>R$}
\label{tab:0-2}
\begin{tabular}{|c|c|c|c|}
\hline
$\gamma$ & $\cos\beta(D-R,0)$  &$\cos\beta(R+D,\pi)$ & $C_1$\\ \hline
$0<\gamma<D-R$         &$-1$ &$-1$&$0$                                  \\ \hline
$D-R<\gamma<R+D$ & $-(D-R)/\gamma$ & $-1$ & $(R+\gamma-D)/2\gamma$\\ \hline
$\gamma>R+D$ &$-(D-R)/\gamma$ & $-(R+D)/\gamma$ &$R/\gamma$\\ \hline
\end{tabular}
\end{table}

\subsection{Under attractive interaction $\gamma<0$}
\subsubsection{Monopole inside the sphere $D<R$}
The singularities sit at $\theta = \pi/2$; thus, the 1st Chern number
\begin{eqnarray}
C_1 &=& \frac{1}{2}\int_0^\pi {\rm d} \theta \sin\beta \frac{\partial \beta}{\partial \theta} \notag\\
&=&
-\frac{1}{2}\left[\left.\cos\beta(k(\theta),\theta)\right|_{\theta=0}^{\pi/2-0^+}+\left.\cos\beta(k(\theta),\theta)\right|_{\theta=\pi/2+0^+}^{\pi}\right].
\end{eqnarray}
Using \textcolor{black}{Eq.(10)} in the text, we could get Table \ref{tab:3}.
($\cos\beta$ at $\theta=\pi/2$ is given by \textcolor{black}{Eq.(15)} in the text.) 
\begin{table}[htbp]
\centering
\caption{$\cos\beta$ at $\theta=0,\pi/2,\pi$, under attractive interaction $\gamma<0$}
\label{tab:3}
\begin{tabular}{|c|c|c|}
\hline
$\theta$    &$k(\theta)$& $\cos\beta$                        \\ \hline
$0$         &$R-D$ &1                                  \\ \hline
$\pi/2+\epsilon$ & $\sqrt{R^2-D^2}$&$\left\{\begin{array}{cc}
\frac{\sqrt{\gamma^2+D^2-R^2}}{\gamma}\text{sign}(\epsilon) & \gamma^2+D^2>R^2\\
0 & \gamma^2+D^2<R^2
\end{array} \right.$  \\ \hline
$\pi$       &$R+D$& -1                                 \\ \hline
\end{tabular}
\end{table}

Thus, based on \ref{tab:3}, we have
\begin{equation}
C_1(\gamma<0,D<R) = \left\{
        \begin{array}{cc}
          1 & \gamma^2 +D^2 <R^2 \\
          1+ \frac{\sqrt{\gamma^2 + D^2-R^2}}{\gamma} &  \gamma^2 +D^2 >R^2  \\
        \end{array}
      \right..
\label{eqn:C13}   \end{equation}
\subsubsection{Monopole outside the sphere $D>R$}

In this case, $S^2$ does not intersect with the singular disc. We only need to consider $\beta(k_-, \theta = \pi)$ and $\beta(k_+, \theta = \pi)$. Due to \textcolor{black}{Eq. (10)} in the text, $\cos\beta(k_-, \theta = \pi) = \cos\beta(k_+, \theta = \pi)=1$ minimizes the ``Zeeman" energy.
Applying textcolor{black}{Eq.(14)} in the text , we have 
\begin{equation}
C_1(\gamma<0, D>R) = 0. 
\label{eqn:C14}
\end{equation}
These results, Eq. (\ref{eqn:C11})(\ref{eqn:C12})(\ref{eqn:C13})(\ref{eqn:C14}), give the curve $C_1$ in \textcolor{black}{Fig. 3(a)} in the main text. 

\section{The 2nd Chern number in terms of ``spin" texture of the Yang monopole}

Using the fact that $P^2 =P$, and $P+ \overline{P} =1$, we have  $PdP = dP \overline{P}$ and $dP P = \overline{P} dP $. With $\hat{F}_{\mu\nu} = -iP[\partial_{\mu}P, \partial_{\nu}P]$, we have 
$\hat{F}_{\mu\nu}\hat{F}_{\rho\sigma} = -  P[\partial_\mu P,\partial_\nu P] P[\partial_\rho P,\partial_\sigma P]= -P [\partial_\mu P,\partial_\nu P] [\partial_\rho P,\partial_\sigma P]$, and ${\rm Tr}\hat{F}_{\mu\nu}=0$,  (as seen from the representation  $P=(1+\bm \hat{\bf m} \cdot \bm \Gamma)/2$), we have 
\begin{eqnarray}
C_2 &=& -\frac{1}{8\pi^2}\int_{S^4} {\rm d}^{4}k \,\,  \epsilon^{\mu\nu\rho\delta}
 Tr(P\partial_\mu P\partial_\nu P\partial_\rho P\partial_\delta P) \\
 &=& -\frac{1}{8\pi^2}\frac{1}{2^5}\int_{S^4} {\rm d}^{4}k \,\, \epsilon^{\mu\nu\rho\delta}\left[
(1+\hat{m}_a \Gamma_a)
(\partial_\mu \hat{m}_b \Gamma_b)
(\partial_\nu \hat{m}_c \Gamma_c)
(\partial_\rho \hat{m}_d \Gamma_d)
(\partial_\sigma \hat{m}_e \Gamma_e)\right]. 
\end{eqnarray}
It is easy to see that due to anti-symmetry, only the terms with all different Gamma matrices survive. We then have 
${\rm Tr } \left[(1+\hat{m}_a \Gamma_a)(\partial_\mu \hat{m}_b \Gamma_b) 
(\partial_\nu \hat{m}_c \Gamma_c) (\partial_\rho \hat{m}_d \Gamma_d)
(\partial_\delta \hat{m}_e \Gamma_e) \right]$
$=\hat{m}_a \cdot \partial_\mu \hat{m}_b\cdot \partial_\nu \hat{m}_c \cdot\partial_\rho \hat{m}_d \cdot\partial_\delta \hat{m}_e {\rm Tr}\left[\Gamma_a\Gamma_b\Gamma_c\Gamma_d\Gamma_e\right]=(\hat{m}_a  \partial_\mu \hat{m}_b  \partial_\nu \hat{m}_c  \partial_\rho \hat{m}_d \partial_\delta \hat{m}_e ) \epsilon^{abcde}  {\rm Tr}[-\mathbb 1]$, and 
\begin{eqnarray}
C_2 &=& \frac{1}{8\pi^2}\frac{1}{2^3}\int_{S^4} {\rm d}^{4}k
\epsilon^{\mu\nu\rho\delta} \epsilon^{abcde} \hat{m}_a  \partial_\mu \hat{m}_b  \partial_\nu \hat{m}_c  \partial_\rho \hat{m}_d \partial_\delta \hat{m}_e \\
&=& \frac{1}{8\pi^2}\frac{1}{2^3}\int_{S^4} {\rm d}^{4}k \epsilon^{\mu\nu\rho\delta}
 \left|
\begin{array}{ccccc}
 \hat{m}_1 & \hat{m}_2 & \hat{m}_3 & \hat{m}_4 & \hat{m}_5 \\
 \partial_\mu \hat{m}_1 & \partial_\mu \hat{m}_2 & \partial_\mu \hat{m}_3 & \partial_\mu \hat{m}_4 & \partial_\mu \hat{m}_5 \\
\partial_\nu \hat{m}_1 & \partial_\nu \hat{m}_2 & \partial_\nu \hat{m}_3 & \partial_\nu \hat{m}_4 & \partial_\nu \hat{m}_5 \\
\partial_\rho \hat{m}_1 & \partial_\rho \hat{m}_2 & \partial_\rho \hat{m}_3 & \partial_\rho \hat{m}_4 & \partial_\rho \hat{m}_5 \\
\partial_\sigma \hat{m}_1 & \partial_\sigma \hat{m}_2 & \partial_\sigma \hat{m}_3 & \partial_\sigma \hat{m}_4 & \partial_\sigma \hat{m}_5 \\
\end{array}
\right|.
\label{C2a}\end{eqnarray}
The 5D parameter space is parametrized by the vector 
${\bf b}= (b_1, b_2, b_3, b_4, b_5) = (k_x, k_y, k_z \hat{n_x}, k_z \hat{n}_{y}, k_z \hat{n}_{z})$, and 
$\hat{\bf m} = (\hat{m}_1, \hat{m}_2, \hat{m}_3, \hat{m}_4, \hat{m}_5) =
(\hat{\ell}_{x}, \hat{\ell}_{y}, \hat{\ell}_{z}\hat{n}_{x}, 
\hat{\ell}_{z}\hat{n}_{y}, \hat{\ell}_{z}\hat{n}_{z})$; 
where $\hat{\bf n}$ is  a unit vector, and ${\bf b}^2 = {\bf k}^2$. In the text, we have denoted the  the polar angles of 
$\hat{\bf k}$,  $\hat{\bf n}$, and $\bm \hat{\ell}$ as  $(\theta, \phi)$,  $(\theta', \phi')$, and $(\beta, \alpha)$; with 
$\beta=\beta(k , \theta)$ and $\alpha=\phi$.   The function $\beta(k , \theta)$ is given by \textcolor{black}{Eq.(10)} in the text. In terms of these polar angles, we have 
\begin{equation}
C_{2}= \frac{3}{8\pi^2}\int_{S^4} {\rm d}\theta {\rm d}\phi {\rm d}\theta' {\rm d}\phi' 
 \left| \begin{array}{ccccc}
\hat{m}_1 & \hat{m}_2 & \hat{m}_3 & \hat{m}_4 & \hat{m}_5 \\
\partial_{\theta} \hat{m}_1 & \partial_{\theta} \hat{m}_2 & \partial_{\theta} \hat{m}_3 & \partial_{\theta} \hat{m}_4 & \partial_{\theta} \hat{m}_5 \\
\partial_{\phi} \hat{m}_1 & \partial_{\phi} \hat{m}_2 & \partial_{\phi} \hat{m}_3 & \partial_{\phi} \hat{m}_4 & \partial_{\phi} \hat{m}_5 \\
\partial_{\theta'} \hat{m}_1 & \partial_{\theta'} \hat{m}_2 & \partial_{\theta'} \hat{m}_3 & \partial_{\theta'} \hat{m}_4 & \partial_{\theta'} \hat{m}_5 \\
\partial_{\phi'} \hat{m}_1 & \partial_{\phi'} \hat{m}_2 & \partial_{\phi'} \hat{m}_3 & \partial_{\phi'} \hat{m}_4 & \partial_{\phi'} \hat{m}_5 \\
\end{array}\right|.
\label{C2b}\end{equation}
The factor 3 comes the fact that there are $4!$ ways
to distribute $\theta, \phi, \theta', \phi'$ among the four indices $\mu, \nu, \rho, \delta$.  Evaluating the determinant, we obtain 
\begin{equation}
C_2 = \frac{3}{4} \int_{S^4} {\rm d}\theta 
{\rm d}\theta' \,\,  {\rm cos}^2\beta {\rm sin}\beta \frac{ \partial \beta}{\partial \theta}\,\, {\rm sin}\theta' . 
\label{C2c} \end{equation}
The $\phi$ and $\phi'$ integration will gives a factor of $(2\pi)^2$, and the factor 3/4 in Eq.(\ref{C2c}) should be 3/2. 
However, since the $5D$ vector ${\bf b}=(b_1,b_2,b_3,b_4,b_5) = (k_x, k_y, k_z \hat{n}_{x}, k_z \hat{n}_{y}, k_z \hat{n}_{z})$, 
$(k_{x}, k_{y}, k_{z}; \hat{\bf n})$ and $(k_{x}, k_{y}, -k_{z}; - \hat{\bf n})$ are the same point. To make the integral in Eq.(\ref{C2c}) more symmetric, 
we  integrate over the entire 2-sphere of $\hat{\bf k}$ and $\hat{\bf n}$, (hence covering $S^{4}$ twice), and then 
divide the result by a factor of 2. The factor 3/4 in Eq.(\ref{C2c}) means we perform the $\theta$ and $\theta'$ integration over $S^4$ in the aforementioned symmetric fashion, with  $0< \theta, \theta'< 2\pi$. 

We  now calculate $C_2$ on  a 4-sphere $S^4$ with radius $R$, centered at a distance $D$ below the monopole  along $\hat{\bf e}_{5}$. With the monopole located at ${\bf b}=0$, the equation of $S^4$ in ${\bf b}$-space is  $({\bf b} + D\hat{\bf e}_{5})^2=R^2$, i.e.
\begin{equation}
k^2 + 2kD{\rm cos}\theta {\rm cos}\theta' + D^2 = R^2. 
\label{4S} \end{equation}
On the surface $S^4$,  we have $k=k(\theta, \theta')$. Since  $\beta$ in Eq.(\ref{C2c}) is a function of $k$ and $\theta$.  Thus, through the relation $\beta=\beta(k, \theta)$, $\beta$ is a function of $\theta$ and $\theta'$, in the form  $\beta(k(\theta, \theta'), \theta)$.  As a result, the two integrations in (\ref{C2c})  can not be performed independently.

\section{The 2nd Chern number of the stretched Yang monopole}

\subsection{Under repulsive interaction $\gamma>0$} 
\subsubsection{Monopole inside the sphere $D < R$} 
On $S^4$, we have 
\begin{equation}
k(\theta,\theta') = -D \cos \theta \cos\theta' + \sqrt{D^2 \cos^2 \theta \cos^2\theta'-D^2 + R^2}, 
\end{equation}
while $\beta=\beta(k,\theta)$ is a function of $k$ and $\theta$ as we discussed in the text. 
The situation is similar to that in \textcolor{black}{Fig.2(a)} in the text, $\theta$ varies from 0 to $\pi$ as $S^{4}$ is covered. 
So we have 
\begin{equation}
C_2 = -\frac{1}{4}\int_0^\pi d\theta' \sin\theta' [\cos^3 \beta(k(\theta= \pi, \theta'), \theta= \pi)-
\cos^3 \beta(k(\theta=0, \theta'), \theta= 0) ].
\end{equation}
Using  \textcolor{black}{Eq.(11)} in the text for ${\rm cos}(k,\theta)$, we have Table \ref{tab:1}
\begin{table}[htbp]
\centering
\caption{$\cos\beta(k(\theta=0, \theta'), \theta= 0)$ and $\cos\beta(k(\theta=\pi, \theta'), \theta= \pi)$ under repulsive interaction $\gamma>0$}
\label{tab:1}
\begin{tabular}{|c|c|c|}
\hline
$\gamma$         & $\cos\beta(k(\theta=0, \theta'), \theta= 0)$ & $\cos\beta(k(\theta=\pi, \theta'), \theta= \pi)$ \\ \hline
$0<\gamma<R-D$   & $1$                           & $-1$                            \\ \hline
$R-D<\gamma<R+D$ &    $\left\{\begin{array}{cc}
k(\theta=0,\theta')/\gamma & \cos\theta' > \frac{R^2-D^2-\gamma^2}{2D\gamma}\\
1 & \cos\theta' < \frac{R^2-D^2-\gamma^2}{2D\gamma}
\end{array} \right.$                 &    $\left\{\begin{array}{cc}
-k(\theta=\pi,\theta')/\gamma & \cos\theta' < \frac{-R^2+D^2+\gamma^2}{2D\gamma}\\
-1 & \cos\theta' >\frac{-R^2+D^2+\gamma^2}{2D\gamma}
\end{array} \right.$                             \\ \hline
$\gamma>R+D$     & $k(\theta=0,\theta')/\gamma$          & $-k(\theta=\pi,\theta')/\gamma$         \\ \hline
\end{tabular}
\end{table}

From Table \ref{tab:1}, we have 
\begin{equation} 
C_2(\gamma>0,D<R) = \left\{\begin{array}{cc}
        1 & 0<\gamma<R-D \\
        f^+(\gamma,D) & R-D<\gamma<R+D \\
        R^3/\gamma^3 & \gamma>R+D
      \end{array}\right.,
\label{eqn:C21}
\end{equation}
where
\begin{equation}
f^+(\gamma,D) = \frac{(\gamma+R-D)^2 \left(D^2+2DR+2D\gamma-3R^2+6R\gamma-3\gamma^2\right)}{16 \gamma^3 D}.
\label{f} \end{equation}

\subsubsection{Monopole outside the sphere, $D > R$}
\begin{figure}[htbp]
\centering
\includegraphics[width=3in]{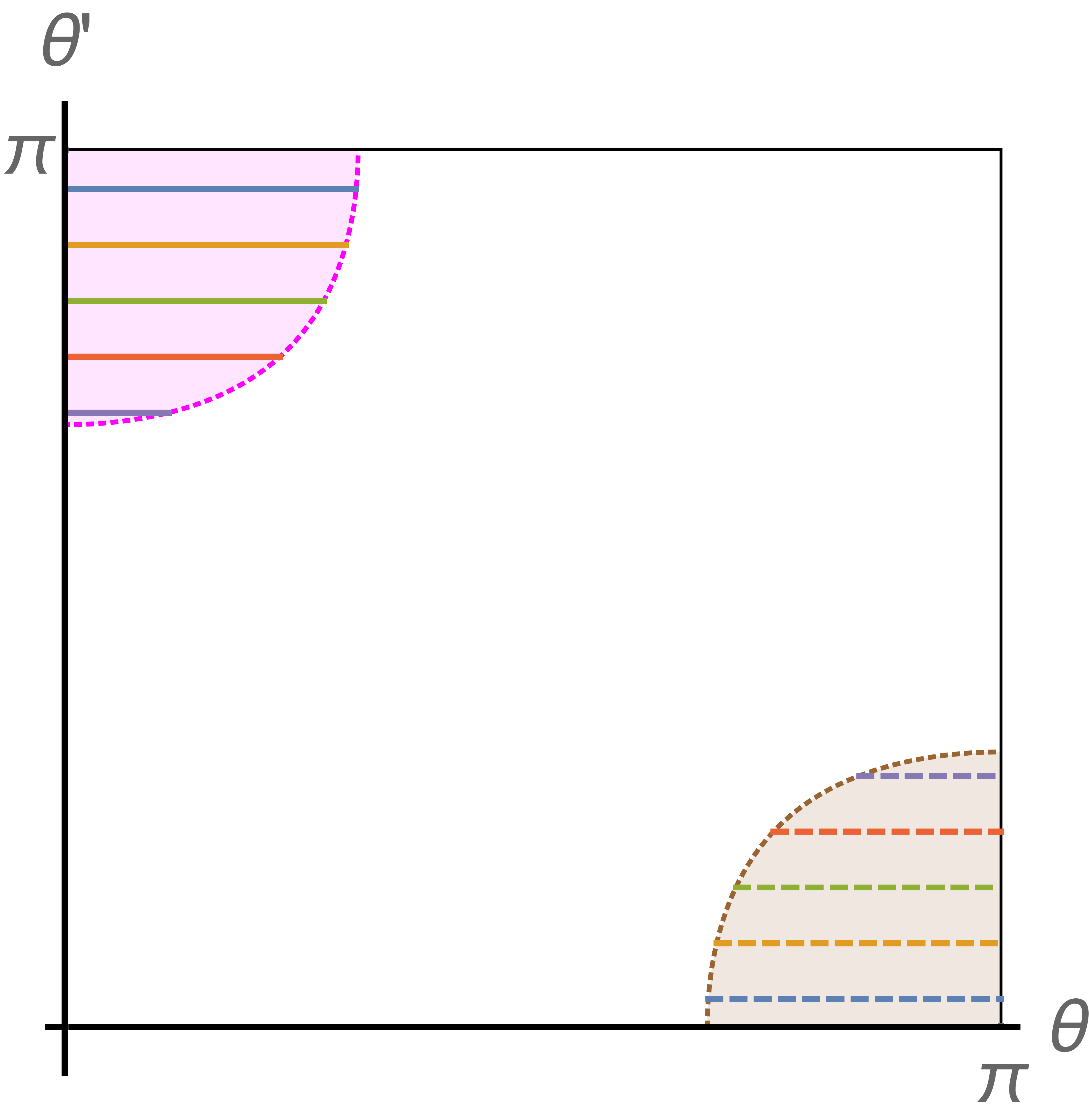}
\caption{The integration region when the monopole is outside $S^4$.The integrals over the brown region (${\rm cos}\theta <0$, ${\rm cos}\theta' >0$) and pink region (${\rm cos}\theta >0$, ${\rm cos}\theta' <0$) give the same result, we can evaluate the integral in one region and multiply the result by $2$. To evaluate $C_{2}$, we hold $\theta'$ fixed and first integrate over $\theta$ along the trajectory shown in the Fig \ref{fig:area}.}
\label{fig:area}
\end{figure}

In this case, there are two branches for $k$ in the $\theta$-$\theta'$ plane, 
\begin{equation}
k_\pm(\theta,\theta') = -D \cos \theta \cos\theta' \pm \sqrt{D^2 \cos^2 \theta \cos^2\theta'-D^2 + R^2},
\end{equation}
with $\theta$ and $\theta'$ restricted to 
\begin{equation}
-1<\cos\theta\cos\theta'<-\sqrt{\frac{D^2-R^2}{D^2}}. 
\end{equation}
The shape of the integration region is shown in Fig.\ref{fig:area}, where the boundary, represented by the dotted line, is given by $\cos\theta\cos\theta'=-\sqrt{\frac{D^2-R^2}{D^2}}$.  
Since the integrals over the region (${\rm cos}\theta <0$, ${\rm cos}\theta' >0$) and (${\rm cos}\theta >0$, ${\rm cos}\theta' <0$) give the same result, we can evaluate the integral in one region and multiply the result by $2$.  

To evaluate $C_{2}$, we hold $\theta'$ fixed and first integrate over $\theta$ along the trajectory shown in the Fig \ref{fig:area}. Noting that there are two branches of $k_{\pm}$ for each  $\theta$, similar to the calculation of $C_{1}$ when the monople is outside the integration surface
as shown in \textcolor{black}{Fig.2(b)} in the text.  We then  have 
\begin{equation}
C_2 = -\frac{1}{2} \int^{\theta^*}_\pi d\theta' \sin\theta' [\cos^3 \beta(k_{+}(\theta=0, \theta'), \theta=0) -\cos^3\beta (k_{-}(\theta=0, \theta'), \theta=0)],
\end{equation}
where
\begin{equation}
\theta^* = \pi - \arccos\sqrt{\frac{D^2-R^2}{D^2}}. 
\end{equation}
Using the expressions of \textcolor{black}{Eq.(11)} in the text for ${\rm cos}(k,\theta)$, we have 
\begin{table}[htbp]
\centering
\caption{$\cos\beta(k_{+}(\theta=0, \theta'), \theta=0)$ and $\cos\beta(k_{-}(\theta=0, \theta'), \theta=0)$ under repulsive interaction $\gamma>0$}
\label{tab:2}
\begin{tabular}{|c|c|c|}
\hline
$\gamma$                    &$\cos\beta(k_{+}(\theta=0, \theta'), \theta=0)$ & $\cos\beta(k_{-}(\theta=0, \theta'), \theta=0)$ \\ \hline
$0<\gamma<D-R$              & $1$                       & $1$                         \\ \hline
$D-R<\gamma<\sqrt{D^2-R^2}$ & $1$                       &        $\left\{\begin{array}{cc}
k_-(\theta=0)/\gamma & \cos\theta' < \frac{R^2-D^2-\gamma^2}{2D\gamma}\\
1 & \cos\theta' > \frac{R^2-D^2-\gamma^2}{2D\gamma}
\end{array} \right.$                     \\ \hline
$\sqrt{D^2-R^2}<\gamma<R+D$ &  $\left\{\begin{array}{cc}
1 & \cos\theta' < \frac{R^2-D^2-\gamma^2}{2D\gamma}\\
k_+(\theta=0)/\gamma & \cos\theta' > \frac{R^2-D^2-\gamma^2}{2D\gamma}
\end{array} \right.$                         & $k_-(\theta=0)/\gamma$      \\ \hline
$\gamma>R+D$                & $k_+(\theta=0)/\gamma$    & $k_-(\theta=0)/\gamma$      \\ \hline
\end{tabular}
\end{table}

From Table \ref{tab:2}, we have 
\begin{equation}
\label{eqn:C22}
C_2(\gamma>0,D>R) = \left\{\begin{array}{cc}
        0 & 0<\gamma<D-R \\
        f^+(\gamma,D) & D-R<g<D+R \\
        R^3/\gamma^3 & g>D+R
      \end{array}\right..
\end{equation}
Here, $f^+(\gamma,D)$ has been defined in Eq.(\ref{f}). 

\subsection{Under attractive interaction $\gamma < 0$}
Due to the orientation of the singular manifold, the calculation of $C_2$ is simpler for $\gamma<0$. 

\subsubsection{Monopole inside the sphere $D<R$}
The singularities sit at $\theta = \pi/2$; thus, the 2nd Chern number
\begin{equation}
C_2 = -\frac{1}{4} \int_0^\pi d\theta'\sin\theta' \left[\left.\cos^3\beta\right|_{\theta=0}^{\pi/2-0^+}+\left.\cos^3\beta\right|_{\theta=\pi/2+0^+}^{\pi}\right].
\end{equation}

As we could treat the Yang monopole in 5D space as 3D monopole with $S^2$ attached to each point, $\beta$ under the Yang monopole case obey the same relation as the Dirac monopole case. Thus, $\beta$ could inherit the result we got in previous section, Table \ref{tab:3}. Based on Table \ref{tab:3}, we have
\begin{equation}
C_2(\gamma<0,D<R) = \left\{
        \begin{array}{cc}
          1 & \gamma^2 +D^2 <R^2 \\
          f^-(\gamma,D) &  \gamma^2 +D^2 >R^2  \\
        \end{array}
      \right..
\label{eqn:C23}
\end{equation}
Here
\begin{equation}
f^-(\gamma,D)=1+ \frac{(\gamma^2 + D^2-R^2)^{3/2}}{\gamma^3}.
\end{equation}

\subsubsection{Monopole outside the sphere $D>R$}
Similarly to the Dirac monopole case, when the monopole is outside the sphere $S^4$, the spherical surface does not intersect with the singular disc.
We only need to consider $\beta(k_-, \theta = 0)$ and $\beta(k_+, \theta = 0)$. 
Noting that $\cos\beta = 1$ minimizes the ``Zeeman" energy at both $(k_-, \theta = 0)$ and $(k_-, \theta = 0)$. Thus, we have 
\begin{equation}
C_2(\gamma<0, D>R) = 0. 
\label{eqn:C24}
\end{equation}
These results, Eq. (\ref{eqn:C21})(\ref{eqn:C22})(\ref{eqn:C23})(\ref{eqn:C24}), give the curve $C_2$ in \textcolor{black}{Fig.3(b)}  in the main text. 

\section{Chern numbers as a function of  interaction $\gamma$ and displacement $D$}

\begin{figure}[htbp]
\centering
\includegraphics[width=3in]{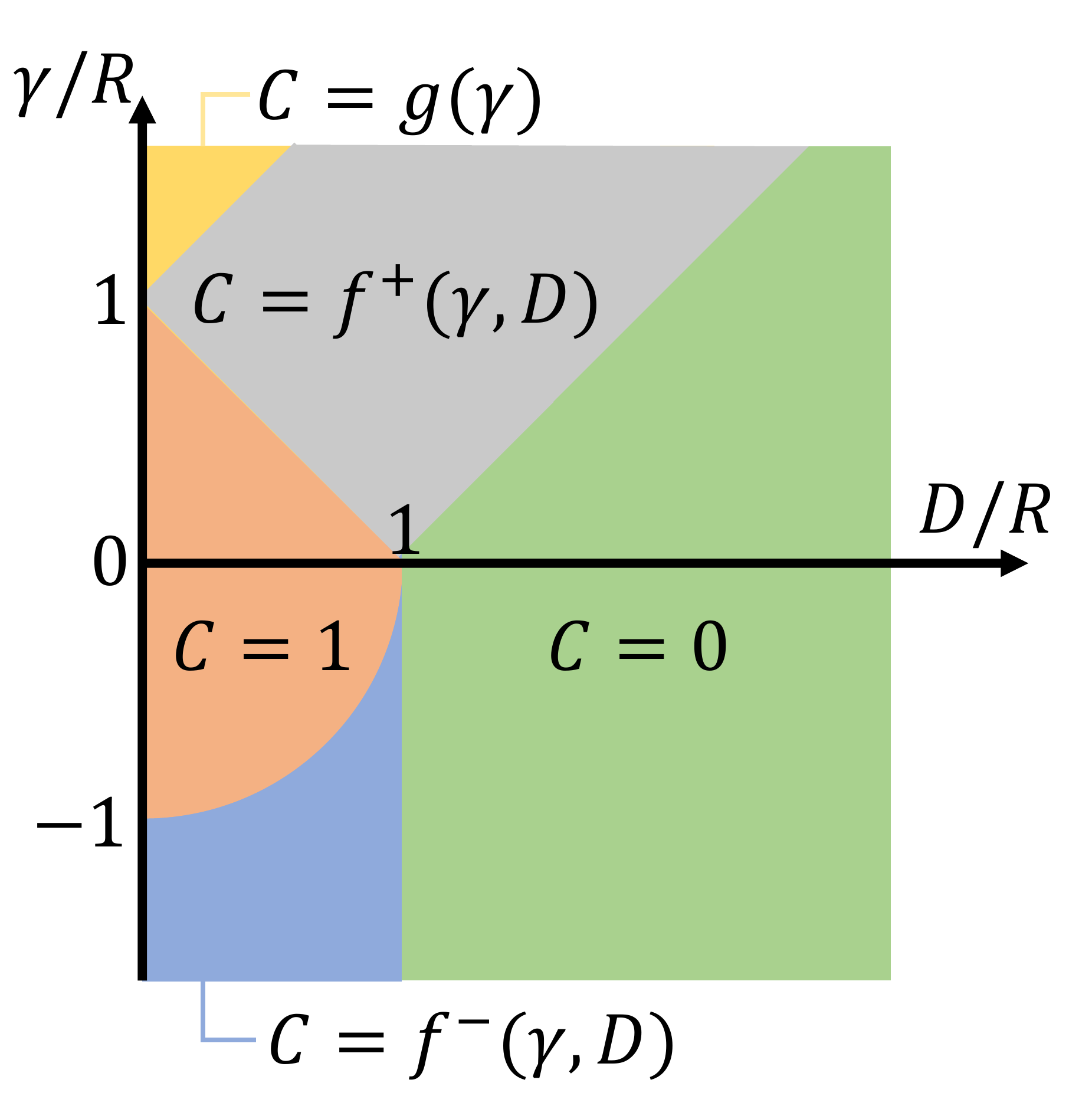}
\caption{1st (or 2nd) Chern number as a function of
interaction strength $\gamma $and displacement $D$ between the
monopole and the center of the integration sphere.}
\label{fig:phase}
\end{figure}

It is also useful to display the value of $C_1$ and $C_2$ as a function of interaction $\gamma$ and displacement $D$ between the monopole and the center of $S^2$ and $S^4$ respectively. The result is shown in  Figure \ref{fig:phase}. The diagram is derived from 
Eq. (\ref{eqn:C11})(\ref{eqn:C12})(\ref{eqn:C13})(\ref{eqn:C14}) for $C_1$; 
and from  Eq. (\ref{eqn:C21})(\ref{eqn:C22})(\ref{eqn:C23})(\ref{eqn:C24}) for $C_2$.  
Figure \ref{fig:phase} shows both  $C_1$ and $C_2$   have the same ``phase-diagram".  Their values in different regions are given by the functions $f^{\pm}$ and $g$. 

For $C_1$, we have 
\begin{eqnarray}
f^+_{(1)}(\gamma,D) &=& \frac{R-D+\gamma}{2\gamma}, \notag\\
f^-_{(1)}(\gamma,D)&=& 1+ \frac{\sqrt{\gamma^2 + D^2-R^2}}{\gamma}, \\
g_{(1)}(\gamma) &=& R/\gamma\notag.
\end{eqnarray}

For $C_2$, we have 
\begin{eqnarray}
f^+_{(2)}(\gamma,D) &=& \frac{(\gamma+R-D)^2 \left(D^2+2DR+2D\gamma-3R^2+6R\gamma-3\gamma^2\right)}{16 \gamma^3 D},\notag\\
f^-_{(2)}(\gamma,D)&=& 1+ \frac{(\gamma^2 + D^2-R^2)^{3/2}}{\gamma^3},\\
g_{(2)}(\gamma) &=& R^3/\gamma^3\notag.
\end{eqnarray}

